\documentclass[12pt]{article}

\def\be{\begin{equation}}
\def\ee{\end{equation}}

\newtheorem{theorem}{Theorem}

\newtheorem{definition}[theorem]{Definition}

\begin{document}

\title{Landscape dynamics, interbasin kinetics \\ and ultrametric diffusion}

\author{S.V.Kozyrev}

\maketitle

\begin{abstract}
We discuss the interbasin kinetics approximation for random walk on
a complex landscape. We show that for a generic landscape the
corresponding model of interbasin kinetics is equivalent to an
ultrametric diffusion, generated by an ultrametric
pseudodifferential operator on the ultrametric space related to the
tree of basins. The simplest example of ultrametric diffusion of
this kind is described by the $p$--adic heat equation.
\end{abstract}

\section{Introduction}

Dynamics of a broad class of complex systems (glasses, clusters,
polymers) is described by a random walk on a complex landscape of
energy \cite{Frauenfelder1}, \cite{Hatom}, \cite{Wales2},
\cite{Wolynes}. Landscape is a real valued function (energy) on a
domain in ${\bf R}^N$. Complex landscape is a function which
possesses many local minima. In particular, models of this kind are
important for description of protein dynamics in the relaxation
approach \cite{Blumenfeld}. Therefore approximations of dynamics on
complex landscapes are important for applications.

We discuss the random walk on the complex energy landscape, given by
the real valued function $U(x)$ on ${\bf R}^N$, with the temperature
$T$ and the inverse temperature $\beta=1/kT$ ($k$ is the Boltzmann
constant). This random walk is defined as follows: the transition
probability rate for transitions between the two neighbor
infinitesimal vicinities $O_1$, $O_2$ of the energy surface will be
proportional to the Boltzmannian factor $\exp(-\beta \Delta U)$,
where $\Delta U$ is the energy difference for the sets $O_1$ and
$O_2$. This formula is valid for transitions which increase energy,
for transitions which decrease energy we put the Boltzmannian factor
equal to one.

For random walk under discussion the system will spend more time in
the low energy areas of the energy landscape. Therefore we get the
following picture --- the system stays in the vicinities of local
minima and performs transitions between local minima through the
energy barriers. For a generic landscape local minima will be
hierarchically clustered with respect to the energy barrier between
the minima.

These arguments suggest the approach of interbasin kinetics, which
is the approximation of a dynamics on a complex landscape, based on
the description of the kinetics of transitions between the groups of
states, called basins. The minimal basins correspond to local minima
of energy, the larger basins are hierarchical unions of smaller
basins.

The postulates of interbasin kinetics:

\medskip

(1) The space of states is separated into basins, basins are
separated into subbasins in a hierarchical way.

\medskip

(2) The activation energy barrier between the two states depends
only on basins, containing these states and does not depend on the
choice of states in these basins.

\medskip

Therefore the transitions between basins are described by the system
of kinetic equations: \be\label{kineq} {d\over d
t}f(i,t)=-\sum_{j}\left[T(i,j)f(i,t)-T(j,i)f(j,t)\right]\nu(j) \ee
Here the indices $i$, $j$ enumerate the states of the system (which
correspond to the minimal basins, or local energy minima),
$T(i,j)\ge 0$ is the probability rate for transitions from $i$ to
$j$, $\nu(j)>0$ are positive numbers (volumes of the basins).

The described constraints of interbasin kinetics on the matrix
$T(i,j)$ imply that this matrix will be a block matrix with a large
number of equal elements. The important example is the Parisi matrix
$T(i,j)$ (used in the theory of spin glasses, see \cite{MPV}).

Various models of interbasin kinetics and hierarchical dynamics were
studied in many papers, see \cite{HoSi}, \cite{StilingerSE},
\cite{Yoshino}, \cite{OS}. In papers \cite{Brekke41},
\cite{Freund94} $p$--adic diffusion was discussed in relation to the
relaxation of spin glasses.

Protein dynamics was studied, with the help of the Mossbauer
spectroscopy by H.Frauenfelder \cite{Frauenfelder1} and
V.I.Goldansky \cite{Gldnsk}. Hierarchical approach to description of
the space of state of a protein was proposed by H.Frauenfelder, see
\cite{Frauenfelder1}.

In \cite{ABKO} it was proposed the approach to describe the
interbasin kinetics models with the help of ultrametric diffusion,
generated by pseudodifferential operators. Namely, the postulates of
interbasin kinetics are put into the form:

\smallskip

(1) = the space of states is ultrametric

(2) = transition probability rate is locally constant

\smallskip

In the simplest case (when $T(i,j)$ is the $p$--adic Parisi matrix
of some simple form, all $\nu(j)$ are equal) the system of equations
of interbasin kinetics takes the form \cite{ABK} of the $p$--adic
heat equation \be\label{padictp} {\partial\over
\partial t}f(x,t)+D_x^{\alpha}f(x,t)=0
\ee Initially this equation was introduced in \cite{VVZ} from purely
mathematical motivations. Here $D_x^{\alpha}$ is the Vladimirov
operator of $p$--adic fractional differentiation with respect to
$x$. This parameter describes the tree of basins for the complex
energy landscape, in the system of equation of interbasin kinetics
(\ref{kineq}) the parameter $x$ corresponds to the index $i$ of the
local minima. For the models of protein dynamics $x$ is the
conformational parameter. $p$--Adic models of interbasin kinetics
were discussed in \cite{ABK}, \cite{ABKO}, \cite{ABO}.

Procedures of construction of the hierarchy of basins and of models
of interbasin kinetics starting from the energy landscape were
studied by Stilinger and Weber \cite{StiWeb1}, \cite{StiWeb2},
Becker and Karplus \cite{BeckerKarplus}. These models were applied
to construction of hierarchy of basins for peptides
\cite{BeckerKarplus} using the data of molecular dynamics. Complex
landscape in this approach is approximated by the disconnectivity
graph and the function of energy barriers on this graph.

In the present paper we construct the equation of ultrametric
diffusion which describes the interbasin kinetics approximation for
the dynamics on a complex landscape of a generic form. This equation
has the form of the following ultrametric pseudodifferential
equation \be\label{to_samoe_ure} {\partial\over \partial
t}f(x,t)+\int_{X}{e^{-\beta E({\rm sup}(x,y))}\over \nu({\rm
sup}(x,y))}\left[e^{\beta E(x)}f(x,t)-e^{\beta
E(y)}f(y,t)\right]d\nu(y)=0 \ee Here $x, y \in X$ lie in the
ultrametric space which describes the tree of basins for the
landscape of energy, $f(x,t)$ is the distribution of occupation. For
a wide class of landscapes the above equations are exactly solvable.
Thus the dynamics on complex landscapes in these cases can be
investigated analytically.  The important example of the above
equation is the $p$--adic heat equation (\ref{padictp}).

The exposition of the present paper is as follows.

In Section 2 we describe the procedure of construction of the tree
of basins and the function of activation energy barriers for a
generic landscape.

In Section 3 we construct the corresponding general model of
interbasin kinetics.

In Section 4 we show the equivalence of the interbasin kinetics
model of Section 3 and the model of ultrametric diffusion on the
space corresponding to the tree of basins.

In Section 5 we discuss the clustering procedure.

In Section 6 we put some material on ultrametric analysis.

\section{Energy landscape and the tree of basins}

Let us describe the procedure which puts into correspondence to an
energy landscape $U$ (a smooth real valued function defined in a
domain (or the configuration space) $M\subset {\bf R}^N$) the tree
of basins, the function on this tree which describes the activation
barriers for a random walk on the landscape, and the measure on the
border of the tree of basins which describes volumes of the
corresponding basins.

Let us consider the set of all local minima of $U$. We assume that
this set is finite. For the local minimum $i$ consider the set
$R(i)$ in the configuration space $M$ (the basin of attraction of
$i$), which contains the points $\xi\in M$, for which:

1) There exists a path (i.e. a continuous curve) in the
configuration space, which connects $\xi$ and $i$, and the function
$U$ does not increase on the path from $\xi$ to $i$.

2) If there exist paths from $\xi$ to several local minima, and the
function $U$ is non increasing along these paths, then the distance
between $\xi$ and $i$ is less or equal than the distances between
$x$ and the other minima. Here the distance between $\xi$ and $i$ is
understood as a distance along the surface of energy, i.e. the
distance between two points of a landscape  is the infimum of
lengths of paths on the energy landscape, which connect the points.

The different $R(i)$, $R(j)$ can intersect on the sets of measure
zero. The union of all $R(i)$ gives the whole configuration space.

Put into correspondence to the basin $R(i)$ the volume $\#(i)$:
$$
\#(i)=\int_{R(i)}dx
$$

\medskip

Let us introduce the following notations.

\medskip

1) Assume that the points $a$, $b$ are connected by the path $S$ in
the configuration space $M$. We say that the point $a$ is separated
from $b$ by the energy barrier $E$ at the path $S$, if the following
supremum over the points $\xi\in S$ is equal to $E$:
$$ {\rm sup}_{\xi\in S}U(\xi)= E
$$

\medskip

2) We say that the points $a$, $b$ in the configuration space are
separated by the energy barrier (or the activation barrier)
$E(a,b)$, if the infimum over the paths $S$ from $a$ to $b$ in the
configuration space of the energy barriers at the path $S$ is equal
to $E(a,b)$:
$$
E(a,b)={\rm inf}_{S}\,{\rm sup}_{\xi\in S}U(\xi)
$$

\medskip

Let $\beta$ be a positive number (the inverse temperature). Let us
introduce on the set of local minima the metric
$$
d(i,j)=e^{-\beta E(i,j)}
$$
For a generic landscape $U$ this metric will satisfy the strong
triangle inequality (i.e. will be an ultrametric).

Let us fix the energy scale --- the increasing sequence of real
numbers $\{E_k\}$, and the corresponding sequence of positive
numbers $D=\{d_k\}$, $d_k=e^{-\beta E_k}$. Consider the
corresponding clustering ${\cal C}_D$ of the set of local minima
with the distance $d(\cdot,\cdot)$ (see the Appendix 1).

The clusters from ${\cal C}_D$ we will also call the basins. Let us
call the directed tree ${\cal T}$ for the clustering ${\cal C}_D$
the disconnectivity graph of the landscape $U$. Using this tree we
build the ultrametric space $X({\cal T})$ (see the Appendix 2). The
points of this space correspond to the local minima of the energy
landscape, the balls (with respect to the ultrametric) correspond to
the basins. One can say that a point $x$ corresponds to some local
minimum $i$ together with the set of inclusions of the corresponding
basins which contain $i$.

On the space $X$ there exists the natural measure $\nu$, such that
the measure $\nu(x)$ of the point $x\in X$ (the space $X$ in the
case under consideration consists of the finite number of points) is
equal to the volume of the basin of attraction of the local minimum
$\#(i)$.

\section{Our ansatz of interbasin kinetics}

Consider the tree of basins for the energy landscape, built with the
help of the procedure of the previous section. Let us construct the
system of equations of interbasin kinetics using the
Arrhenius--Eyring formula, which gives the approximation for the
velocity constant of reaction in chemical kinetics:
$$
\kappa=A\exp(-\beta\Delta F)
$$
where $\kappa$ is the velocity constant of reaction, $\Delta F$ is
the free energy of activation, $A$ is some constant, $\beta$ is the
inverse temperature. Let us remind that the free energy of the group
of states (with the same energy) is defined as
$$
F=E-\theta S
$$
where $E$ is the energy of the group of states, $\theta=\beta^{-1}$
is the temperature, $S$ is the entropy (logarithm of the number of
states in the group).

We consider the system of equations of interbasin kinetics of the
form
$$
{d g(i,t)\over dt}=-\sum_{j\ne i}\left[e^{\beta(F(i)-G({\rm
sup}(i,j)))}C(i,j)g(i,t)-e^{\beta(F(j)-G({\rm
sup}(i,j)))}C(j,i)g(j,t)\right]
$$
Here $i$, $j$ are minimal basins (which correspond to local minima),
$g(i)$ is the occupation of the minimal basin $i$, $F(i)$ is the
free energy of the basin $i$, ${\rm sup}(i,j)$ is the minimal
superbasin which contains both the basins $i$ and $j$, $G({\rm
sup}(i,j))$ is the free energy of the transition state for
transitions between $i$ and $j$.

This system of equations is based on the Eyring formula and the
assumption that the transition state for transitions between $i$ and
$j$ is defined by the superbasin ${\rm sup}(i,j)$).

We choose the coefficients $C(i,j)$ to be positive and symmetric. In
this case the above system of kinetic equations satisfies the
conditions of detailed balance. These coefficients describe the
modification of the Eyring formula on the case of transitions
between the groups of states with the unique intermediate transition
state. Choice of the coefficients $C(i,j)$ fixes the model of
interbasin kinetics. We propose the following ansatz for the
coefficients: \be\label{cond1} C(i,j)={\#(i)\#(j)\over \#^2({\rm
sup}(i,j))} \ee Here $\#(i)$ is the number of states in the basin
$i$ (i.e. the volume of this basin). This choice satisfies the
scaling conditions
--- the coefficients $C(i,j)$ do not change with dilatations of the landscape.

With this choice of the coefficients the system of equations of
interbasin kinetics takes the form
$$
{d f(i,t)\over dt}=-\sum_{j\ne i}{e^{-\beta G({\rm sup}(i,j))}\over
\#^2({\rm sup}(i,j)) }\left[e^{\beta E(i)}f(i,t)-e^{\beta
E(j)}f(j,t)\right]\#(j)
$$
Here $f(i)=g(i)/\#(i)$ is the density of occupation of the basin
$i$, $E(i)$ is the energy of the basin $i$ (i.e. $e^{\beta
F(i)}=e^{\beta E}/\#(i)$). We choose the volumes of the transition
states for basins ${\rm sup}(i,j)$ to be proportional to the volumes
of these basins: \be\label{cond2} e^{S({\rm sup}(i,j))} \sim \#({\rm
sup}(i,j) \ee where $S({\rm sup}(i,j))$ is the entropy of the
transition state. With this choice of entropy for transition states
(we ignore the corresponding coefficient of proportionality) the
system of equations of interbasin kinetics takes the form
\be\label{systemIBM} {d f(i,t)\over dt}=-\sum_{j\ne i}{e^{-\beta
E({\rm sup}(i,j))}\over \#({\rm sup}(i,j)) }\left[e^{\beta
E(i)}f(i,t)-e^{\beta E(j)}f(j,t)\right]\#(j) \ee Here $E({\rm
sup}(i,j))$ is the energy of the transition state for the basin
${\rm sup}(i,j)$, and this value coincides with the energy used in
the clustering procedure of construction of the tree of basins.

Therefore the introduced here ansatz of interbasin kinetics is based
on the clustering procedure of construction of the tree of basins,
the Arrhenius--Eyring formula and conditions (\ref{cond1}),
(\ref{cond2}), and generates the system of equations
(\ref{systemIBM}).

\section{Ultrametric diffusion}

In the present section we show that the system of equations of
interbasin kinetics is equivalent to the dynamics on the ultrametric
space $X$ corresponding to the tree of basins.

We have the following theorem.

\begin{theorem}\label{ibm_ud}\qquad {\sl
The system of equations of interbasin kinetics (\ref{systemIBM}) is
equivalent to the ultrametric pseudodifferential equation
\be\label{system_ultra} {\partial\over \partial
t}f(x,t)+\int_{X}{e^{-\beta E({\rm sup}(x,y))}\over \nu({\rm
sup}(x,y))}\left[e^{\beta E(x)}f(x,t)-e^{\beta
E(y)}f(y,t)\right]d\nu(y)=0 \ee where the ultrametric space $X$
corresponds to the tree of basins of the energy landscape, the
points $x$ of the ultrametric space correspond to the minimal basins
$i$ (basins of attraction of local minima), the measure $\nu$
describes volumes of the basins (i.e. for the minimal basin $i$
corresponding to the minimal ball $x$ we have $\nu(x)=\#(i)$).  }
\end{theorem}

We will not restrict the consideration of the dynamics on energy
landscapes to ultrametric spaces containing finite number of points,
but instead we will consider the general case of equations of the
form (\ref{system_ultra}). Finite trees of basins are obtained
because we consider smooth energy landscapes. In reality energy
landscapes can be complex and rugged. For a rugged energy landscape
the described procedure is not directly applicable. Instead we can
consider the inductive limit of directed trees and related spaces of
functions. We investigate the pseudodifferential equation of the
form (\ref{system_ultra}) on the ultrametric space corresponding to
the limiting infinite tree, and interpret this equation as
describing dynamics on a rugged landscape.

\bigskip

\noindent{\bf Example}\qquad Consider the case when $X=Q_p$, the
measure $\nu$ is the Haar measure $\mu$, and the activation energy
is chosen as follows:
$$
E(|x-y|_p)=k\ln |x-y|_p,\qquad k>0,
$$
The potential of the minimal basins (point in $Q_p$) is equal to
zero. In the notation $|x-y|_p=p^{\gamma}$, the activation energy is
linear with respect to $\gamma$. We get for the transition
probability rate the expression
$$
{e^{-\beta E({\rm sup}(x,y))}\over \nu({\rm sup}(x,y))}={e^{-\beta
k\ln |x-y|_p}\over |x-y|_p}={1\over |x-y|_p^{1+\beta k}}
$$
Equation of interbasin kinetics takes the form of the $p$--adic heat
equation
$$
{\partial\over \partial t}f(x,t)+D^{\alpha}_{x}f(x,t)=0
$$
where the parameter $\alpha$ of the Vladimirov operator of the
$p$--adic fractional differentiation
$$
D^{\alpha}_{x}f(x,t)=\Gamma_p^{-1}(-\alpha)\int_{Q_p}{f(x,t)-f(y,t)\over
|x-y|_p^{1+\alpha}}d\mu(x)
$$
is proportional to the inverse temperature: $\alpha=\beta k$.

\bigskip

\noindent{\bf Remark}\qquad Cauchy problem for the $p$--adic heat
equation is exactly solvable. Analogously, Cauchy problem for
equation(\ref{system_ultra}) is exactly solvable (with the help of
the ultrametric wavelet transform) if the energies of local minima
are equal: $E(x)={\rm const}$. Therefore in the interbasin kinetics
approximation the dynamics for a wide class of complex energy
landscapes possesses analytical investigation.

\bigskip

\noindent{\bf Example: Mb--CO rebinding}\qquad One of the most
important applications of the dynamics on energy landscapes  and
interbasin kinetics is the application to conformational dynamics of
proteins. In this case the ultrametric parameter $x$ describes the
conformational coordinate for the protein.

In paper \cite{ABKO} it was shown that the obtained with the help of
$p$--adic methods results on protein dynamics coincide with the data
of spectroscopic experiments for Mb--CO rebinding. Mb--CO rebinding
is a fundamental model in the physics of proteins and plays the role
of ''the hydrogen atom of biology'' \cite{Hatom}.

Let us describe the approach of \cite{ABKO}. Myoglobin can bind CO
only when myoglobin is in some particular subset of the space of
conformations (when the path to the active center of the molecule is
opened). Consider the model of Mb--CO rebinding described by the
equation of interbasin kinetics \be\label{mioglobinCO}
\left[{\partial\over
\partial t}+D_x^{\alpha}+\Omega(|x|_p)\right]f(x,t)=0
\ee Here $\alpha$ is proportional to the inverse temperature
$\beta$, the conformational coordinate is parameterized by the field
of $p$--adic mumbers. The function $f(x,t)$ is the density of
occupation of the space of conformations for molecules of myoglobin
(not bound to CO). The Mb--CO binding takes place on the subset of
the space of conformations described by the unit ball in $Q_p$.

Equation (\ref{mioglobinCO}) is a model of ultrametric diffusion
with a sink.

\bigskip

\noindent{\bf Remark}\qquad In the model of Mb--CO rebinding
(\ref{mioglobinCO}) we get the generator of diffusion with a sink in
the form of the $p$--adic Schrodinger operator
$$
D_x^{\alpha}+\Omega(|x|_p).
$$
The term with positive potential describes a sink (negative
potential will describe a source). The Vladimirov operator
$D_x^{\alpha}$ plays the role of a $p$--adic Laplacian.

The operator in the RHS of the equation (\ref{system_ultra}) have
the form of the product of operators
$$
Df(x)=\int_{X}{e^{-\beta E({\rm sup}(x,y))}\over \nu({\rm
sup}(x,y))}\left[e^{\beta E(x)}f(x)-e^{\beta
E(y)}f(y)\right]d\nu(y)=TX f(x),
$$
where
$$
Tf(x)=\int_{X}{e^{-\beta E({\rm sup}(x,y))}\over \nu({\rm
sup}(x,y))}\left[f(x)-f(y)\right]d\nu(y),
$$
is the ultrametric pseudodifferential operator and
$$
Xf(x)=e^{\beta E(x)}f(x)
$$
is the operator of multiplication by the exponent of the potential.

Therefore in applications of ultrametric analysis to models of
interbasin kinetics we get the Schrodinger operator (a sum of a
pseudodifferential operator and an operator of multiplication by a
function), and a product of a pseudodifferential operator and an
operator of multiplication by a positive function.

\section{Appendix 1: Clustering}

In the present section we discuss the clustering procedure for
metric spaces. Denote $(M,\rho)$ the metric space $M$ with metric
$\rho$.

\begin{definition}{\sl
A sequence of points $a=x_0,x_1,\dots,x_{n-1},x_n=b$ in the metric
space $(M,\rho)$ is called an $\varepsilon$--chain connecting $a$
and $b$, if $\rho(x_k,x_{k+1})\le\varepsilon$ for all $0\le k <n$.
If there exists an $\varepsilon$--chain connecting $a$ and $b$, we
say that $a$ and $b$ are $\varepsilon$--connected. }
\end{definition}

In an ultrametric space any two points  $a$ è $b$ are not
$\varepsilon$--connected for $\varepsilon<\rho(a,b)$.

\begin{definition}\label{chain_distance}{\sl
Let $(M,\rho)$ be an arbitrary metric space. Let us define the chain
distance $$d(a,b)={\rm inf }\,(\varepsilon: a,b\quad\hbox{are}\quad
\varepsilon\hbox{--connected}).$$}
\end{definition}

The chain distance $d(a,b)$ between the points $a$ and $b$ satisfies
all the properties of ultrametric except for nondegeneracy, i.e.
$$
d(a,b)=d(b,a)\quad\forall a,b,
$$
$$
d(a,b)\le {\rm max}\, \left(d(a,c),d(c,b)\right)\quad\forall a,b,c,
$$
but it is possible that $d(a,b)=0$ for some $a\ne b$.

If the space $M$ is ultrametric (i.e. $\rho$ satisfies the strong
triangle inequality), then the chain distance $d(\cdot,\cdot)$ will
coincide with the ultrametric $\rho(\cdot,\cdot)$.

\begin{definition}\label{cluster}
{\sl Let us call the cluster $C(i,R)$ in the metric space $(M,\rho)$
the ball with respect to the chain distance with the center in $i$
and the radius $R$, i.e. the set $\{j\in M: d(i,j)\le R \}$. The
clustering of the space $M$ is the set of clusters in $M$, such that
any element of $M$ lies in some cluster. }
\end{definition}

By this definition the set of clusterings is partially ordered:
assume we have two clusterings ${\cal A}$ and ${\cal B}$ of the set
$S$, then ${\cal A}>{\cal B}$, if all clusters of ${\cal B}$ are
subsets of clusters of ${\cal A}$.

Since the chain distance satisfies the strong triangle inequality,
any clustering ${\cal C}$ generates a directed tree of clusters
${\cal T}={\cal T}[M]$ and an ultrametric on this tree (the chain
distance between clusters). Then using the standard procedure (see
the Appendix 2) we construct the ultrametric space $X=X({\cal T})$
(the chain space of the clustering ${\cal C}$), which can be
identified with the border of the tree ${\cal T}$. Clusters in the
metric space $M$ correspond to balls in the ultrametric space $X$.

\bigskip

\noindent {\bf Example}\qquad Consider the important example of
clustering. Let $D=\{d_i\}$ be a finite or countable set of positive
numbers without positive accumulation points. Consider the
clustering ${\cal C}_D$ of the metric space $(M,\rho)$ which
contains all clusters of chain radii $d_i\in D$ and arbitrary
centers.

\section{Appendix 2: Ultrametric analysis}

In this Section we discuss some results on ultrametric analysis,
which can be found in  \cite{Izv}, \cite{ACHA}, \cite{MathSbornik}.

\begin{definition}{\sl
An ultrametric space is a metric space with the ultrametric $d(x,y)$
(where $d(x,y)$ is called the distance between $x$ and $y$), i.e. a
function of two variables, satisfying the properties of positivity
and non degeneracy
$$
d(x,y)\ge 0,\qquad d(x,y)=0\quad \Longrightarrow\quad x=y;
$$
symmetricity
$$
d(x,y)=d(y,x);
$$
and the strong triangle inequality
$$
d(x,y)\le{\rm max }(d(x,z),d(y,z)),\qquad \forall x,y,z.
$$
}
\end{definition}

We say that an ultrametric space $X$ is regular, if this space
satisfies the following properties:

\medskip

1) The set of all the balls of nonzero diameter in $X$ is finite or
countable;

\medskip

2) For any decreasing sequence of balls $\{D^{(k)}\}$,
$D^{(k)}\supset D^{(k+1)}$, the diameters of the balls tend to zero;

\medskip

3) Any ball of non--zero diameter is a finite union of maximal
subballs.

\bigskip

Ultrametric spaces are dual to directed trees. Below we describe
some part of the duality construction.

For a regular ultrametric space $X$ consider the set ${\cal T}(X)$,
which contains all the balls in $X$ of nonzero diameters, and the
balls of zero diameter which are maximal subbals in balls of nonzero
diameters. This set possesses a natural structure of a directed
tree. Two vertices $I$ and $J$ in ${\cal T}(X)$ are connected by an
edge if the corresponding balls are ordered by inclusion, say
$I\supset J$ (i.e. one of the balls contain the other), and there
are no intermediate balls between $I$ and $J$.

The partial order in ${\cal T}(X)$ is defined by inclusion of balls,
this partial order is a direction. We recall that a partially
ordered set is a directed set (and a partial order is a direction),
if for any pair of elements there exists the unique supremum with
respect to the partial order.

On the directed tree ${\cal T}(X)$ we have the natural increasing
positive function which puts into correspondence to any vertex the
diameter of the corresponding ball.

Assume now we have a directed tree ${\cal T}$ with the positive
increasing function $F$ on this tree. Then we define the ultrametric
on the set of vertices of the tree as follows: $d(I,J)=F({\rm
sup}(I,J))$ where ${\rm sup}(I,J)$ is the supremum of vertices $I$,
$J$ with respect to the direction.

Then we take completion of the set of vertices with respect to the
defined ultrametric and eliminate from the completion all the inner
points of the tree (a vertex of the tree is inner if it does not
belong to the border of the tree). We denote the obtained space
$X({\cal T})$, this space is ultrametric.

An ultrametric pseudodifferential operator is defined in the
following way. Consider a $\sigma$--additive Borel measure $\nu$
with countable or finite basis on a regular ultrametric space $X$.
Consider the pseudodifferential operator
$$
Tf(x)=\int T{({\rm sup}(x,y))}(f(x)-f(y))d\nu(y)
$$
Here $T{(I)}$ is some complex valued function on the tree ${\cal T}(
X)$. The supremum
$$
{\rm sup}(x,y)=I
$$
of the points $x,y\in X$ is the minimal ball $I$ in $X$, containing
both points.

\bigskip\bigskip

\noindent{\bf Acknowledgments}\qquad The author would like to thank
V.A.Avetisov, A.Kh.Bikulov, I.V.Volovich, V.S.Vla\-di\-mi\-rov and
A.Yu.Khrennikov for fruitful discussions and valuable comments. He
gratefully acknowledges being partially supported by the grant DFG
Project 436 RUS 113/809/0-1, by the grants of The Russian Foundation
for Basic Research  RFFI 05-01-04002-NNIO-a and RFFI 05-01-00884-a,
by the grant of the President of Russian Federation for the support
of scientific schools NSh 6705.2006.1 and by the Program of the
Department of Mathematics of Russian Academy of Science ''Modern
problems of theoretical mathematics''.

\end{document}